\documentclass[fleqn,twoside]{article}
\usepackage{espcrc2}
\usepackage{amsmath}
\usepackage{graphicx}
\usepackage[figuresright]{rotating}

\def\nutau{$\nu_{\tau}$}

\def\gtrsim{\mathrel{\hbox{\rlap{\hbox{\lower4pt\hbox{$\sim$}}}\hbox{$>$}}}}
\def\lesssim{\mathrel{\hbox{\rlap{\hbox{\lower4pt\hbox{$\sim$}}}\hbox{$<$}}}}
\title{Muon and Gamma Bundles tracing  Up-going  Tau Neutrino  Astronomy  }
\author{D.Fargion \address[MCSD]{Physics Department Rome Univeristy\\
INFN,Rome, Italy}, M.De Santis, P.G.De Sanctis Lucentini, M.
Grossi }

\begin{document}

\begin{abstract}

  Up-going and Horizontal Tau Air-Showers, UpTaus and HorTaus,  may trace Ultra High Energy Neutrino
  Tau Earth Skimming at the edge of the horizons. Their secondaries ($\mu^\pm$ and $\gamma$ bundles with $e^\pm$
   pair flashes)
  might trace their nature over UHECR secondaries in horizontal showers. Indeed the atmosphere act as a perfect
   amplifier as well as  a  filter
  for showers: down-ward and horizontal $\mu$ bundles may still
  be originated by far  Ultra High Energy Cosmic Rays skimming the
  terrestrial  atmosphere but their rich gamma component will be exponentially suppressed.
  At large zenith angles after crossing a large slant depth ($X_{max} > 3 \times 10^3$ g cm$^{-2}$)
  the number of $\mu^\pm$ and secondary $\gamma$'s
  (produced by the $e^\pm$ pair from $\mu$ decay in flight) is
  comparable.
  On the other hand, up-ward muon bundles from UpTaus and HorTaus may arise within a young shower with
  a larger gamma-muon ratio ($\sim 10^2$), leaving its characteristic imprint.
  We estimate the UpTaus and HorTaus rate from
  the Earth and we evaluate the consequent event rate of
  $\mu^{\pm}$, $e^\pm$ and $\gamma$ bundles. We show
  that such events even for minimal GZK neutrino fluxes could be detected by scintillator arrays
   placed on  mountains at  $1-5$ km  and pointing to the horizon. The required array
areas  are within tens-hundreds of square meters. An optimal
structure is an array of  crown-like twin detectors facing the
horizons.

   We argue   that such detectors will be able to detect both muonic bundles
   at a minimal average flux of $ 10^{-11}$ cm$^{-2}$ s$^{-1}$
   sr$^{-1}$ and electromagnetic particles ($\gamma$, $e^\pm$) at
$ 3 \times 10^{-9}$ cm$^{-2}$ s$^{-1}$
   sr$^{-1}$, a few times each year, even for the minimal GZK $\nu$ flux.

\end{abstract}





\maketitle

\section{Introduction}

\begin{figure}[t]  
\includegraphics[height=40mm,width=70mm]{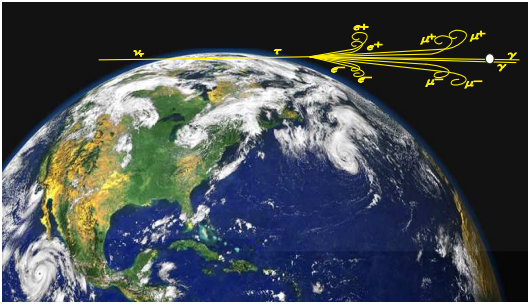}
\caption{Horizontal Upward Tau Air-Shower (HorTauS) originated by
UHE neutrino skimming the Earth: fan-shaped jets arise because of
the geo-magnetic bending of charged particles at high quota ($\sim
23-40$ km). The shower signature may be observable by EUSO just
above the horizon. Because of the Earth
opacity most of the UpTau events at angles $%
\protect\theta > 45-50^o$ will not be observable, since they will
not be contained within its current field of view (FOV).
} \label{fig3}
\end{figure}

\begin{figure}[h]
\includegraphics[width=55mm,angle=270]{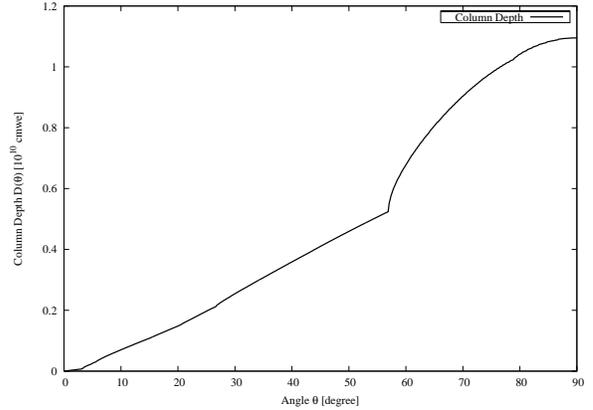}
\caption{Column depth as a function of the incoming angle having
assumed the multi-layers structure  given by the Earth Preliminary
Model (see Fargion et al.2004).} \label{Dteta}
\end{figure}

The study of ultrahigh energy upward and horizontal $\tau$ air
showers produced by $\tau$ neutrino interactions within the Earth
crust has been considered in recent years as an alternative way to
detect high energy neutrinos. The $\nu_\tau$ UHE sources are more  common $\nu_\mu$ (or $\nu_e$) whose tiny masses ($\Delta m_\nu > 5\cdot 10^{-2}$ eV) and flavour mixing would lead to a $\nu_\tau$ appearence.
 The results are often difficult to
interpret. The problem of $\tau$ neutrinos crossing the Earth is
indeed quite complicated because of the complex terrestrial
neutrino opacity at different energies and angles of arrival. In
addition, several factors have to be taken into account, such as
the amount of energy transferred in the $\nu_{\tau}$ - $\tau$
lepton conversion, as well as the $\tau$ energy losses and
interaction lengths at different energies and materials. This makes
the estimate of the links between the input neutrino - output
$\tau$ air shower very difficult. Such a prediction is further
complicated by the existence of a long list of theoretical models
for the incoming neutrino fluxes (GZK neutrinos, Z-burst model
flux, $E^{-2}$ flat spectra, AGN neutrinos, topological defects).
Many authors have investigated this $\nu_\tau$ signature by their convolution of flux models and the Earth new opacity; however the results
are varied, often in contradiction among themselves, and the
expected rates may
range over a few order of magnitude (Fargion, Aiello, \& Conversano
1999; Fargion 2002; Bertou et al. 2002; Feng et al. 2002; Bottai \&
Giurgola 2003, hereafter BG03; Tseng et al. 2003; Bugaev,
Montaruli, \& Sokalski 2004; Fargion et al. 2004;  Jones et al.
2004; Yoshida et al. 2004). So far, the majority of the current
studies on this topic is based on Monte-Carlo simulations assuming
a particular model of the incoming neutrino flux. Most of the
authors focus on the UpTaus tracks in underground detectors.

To face such a complex problem, we considered  the simplest
approach: first we  disentangle the incoming
neutrino flux from the consequent $\tau$ air-shower physics;
therefore, to establish the $\tau$ production rate we introduce
an effective volume and mass for Earth-skimming $\tau$'s, which
is independent on any incoming neutrino flux model. This volume
describes a strip within the Earth crust where
neutrino/antineutrino-nucleon, $\nu_{\tau} (\bar{\nu}_{\tau })
-N$, interactions may produce emerging $\tau^{-},\tau^{+}$
leptons which then shower in the atmosphere outside the Earth.

We present a very simple analytical and numerical derivation (as
well as its more sophisticated extensions) which takes into
account, for any incoming angle, the main processes related to the
neutrinos and $\tau$ leptons propagation and the $\tau$ energy
losses within the Earth crust. Our numerical results are
constrained by upper and lower bounds derived in simple
approximations (see Fargion et al. 2004 for details). The effective
volumes and masses will be more severely reduced at high energy
because we are interested in the successful development of the
$\tau$ air-shower. Therefore we included as a further constraint
the role of the air dilution at high altitude, where $\tau$ decay
and the consequent air-shower may (or may not) take place.


We showed (Fargion et al. 2004) that our results give an estimate
of the $\tau$ air-shower event rates that exceeds earliest studies
but they were comparable or even below more recent predictions
(Yoshida et al. 2004). Secondly, we  point out here, for the first time, that the
consequent $\mu^{\pm}$, $e^{\pm}$, $\gamma$ signature of HorTaus
largely differs from that of horizontal UHECR backgrounds. Finally
we introduce in the calculation of the number of events an
additional suppression factor related to the altitude at which air
showers are observed. This guarantees the optimal extension and the
largest flux for the shower to be detected at each observational $h$ altitude.


\begin{figure}[t] 
\includegraphics[width=55mm,angle=270]{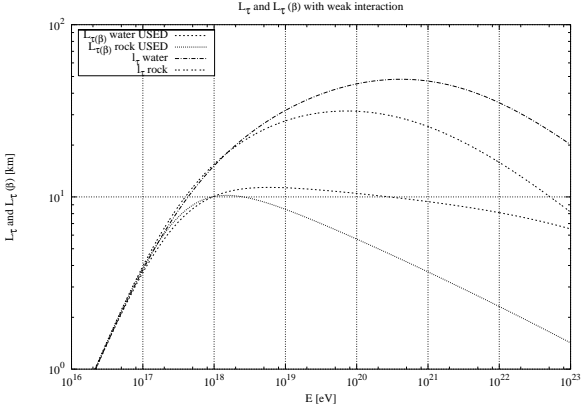} \caption{Comparison between $L_{\tau (\beta)}$
and $l_{\tau}$ for rock and water.  As one can see from the
picture, $L_{\tau (\beta)}$ is shorter than $l_{\tau}$ at energies
above $10^{17}$ eV, thus it
 corresponds to a smaller effective volume where
$\protect\tau$'s are produced while keeping most of the primary
neutrino energy. The energy   label on the x axis refers to the
newly born tau for $L_{\tau(\beta)}$: $E_{\nu_{\tau}}$}
\label{l_tau}
\end{figure}

\section{Effective volume and masses for HorTaus}

To calculate the effective volume we assume that the neutrino
traversing the Earth is transformed in a tau lepton at a depth $x$,
after having travelled for a distance ($D(\theta) - x$). The column
depth $D(\theta)$ defined as $\int \rho(r) dl$, the integral of the
density $\rho(r)$ of the Earth along the neutrino path at a given
angle $\theta$ is shown in Fig. \ref{Dteta}. The angle $\theta$ is
included between the neutrino arrival direction and the tangent
plane to the earth at the observer location ($\theta = 0^\circ$
corresponds to a beam of neutrinos tangential to the earth's
surface) and it is complementary to the nadir angle at the same
location. The probability for the neutrino with energy $E_{\nu}$ to
survive until a distance ($D(\theta) - x$) is $e^{-(D(\theta) -
x)/L_{\nu}}$, while the probability for the tau to exit the Earth
is $e^{- x/l_{\tau}}$. On the other hand, as we will show in the
next section, the probability for the outcoming $\tau$ to emerge
from the Earth keeping its primary energy $E_{\tau_i}$ is $e^{-
x/L_{\tau (\beta)}}$ (where $e^{- x/L_{\tau (\beta)}}$ $\ll$ $e^{-
x/l_{\tau}}$ at energy $E_{\tau} > 3 \times 10^{17}$ eV). By the
interaction length $L_{\nu}$ we mean the characteristic length for
neutrino interaction; as we know its value may be associated to the
inverse of the total
cross-section  $\sigma_{Tot}= \sigma_{CC} + \sigma_{NC}$, including
both charged and neutral current interactions. It is possible to
show that using the $\sigma_{CC}$ in the $e^{-(D(\theta) -
x)/L_{\nu}}$ factor includes most of the ${\nu}_{\tau}$
regeneration along the neutrino trajectory making simpler the
mathematical approach.

Indeed the use of the total cross-section in the opacity factor
above  must be corrected  by the multi-scattering events (a neutral
current interaction first followed by a charged current one later);
these additional $relay$ events ("regenerated taus") are summarized
by the less suppressing $\sigma_{CC}$ factor in the $e^{-(D(\theta)
- x)/L_{\nu CC }}$ opacity term. The only difference between the
real case and our very accurate approximation is that we are
neglecting a marginal energy degradation (by a factor $0.8$) for
only those "regenerated taus" which experienced  a previous neutral
current scattering. We also neglected all the charged current
events deep inside the Earth whose tau lepton birth and propagation
is source of lower energy tau neutrinos  which pile up at energy
$10^{17}$ eV  and below ; such a $\nu_{\tau}$ regeneration (absent
in $\nu_{\mu}$ or $\nu_{e}$ cases)  may be here neglected because
of its marginal role in the  range of energy ($ > 10^{17}$ eV)  we
are interested in. Therefore our estimates give a lower bound for
the detection of $\tau$ air showers from UHE $\nu_{\tau}$ skimming
the Earth. The rarer short range decay of UHE $\tau$ inside the
Earth, with a marginal energy loss, may be an additional source of
"UHE regenerated" $\nu_{\tau}$ even at energy $10^{18}$-$ 10^{19}$
eV; their additional contribution (here neglected) will be
discussed elsewhere.

\begin{figure}[t] 
\includegraphics[width=52mm,angle=270]{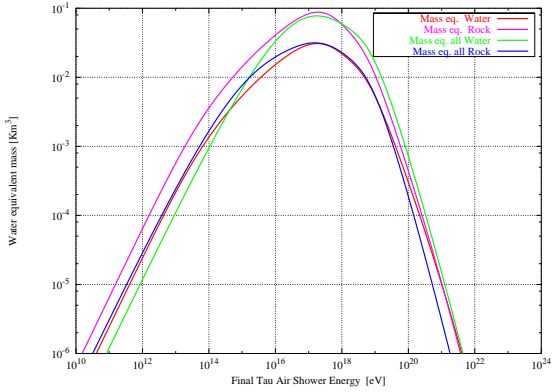}
\caption{Effective mass for UpTaus and HorTaus per km square unit
area including the suppression factor due to the finite extension
of the Earth's atmosphere (an horizontal length  of $600$ km). The
curves obtained (the red line for an Earth outer layer made of
water and the pink line for the rock) are compared with a
simplified model of the Earth, considered as an homogeneous sphere
of water (green line) and rock (blue line). Here we used the
interaction length $L_{\protect\tau (\protect\beta)}$ and the
volume is expressed as a function of the final tau energy. Note
that above $10^{-2}$ km$^2$ the effective mass-volume in the energy
range 3 $\times$ 10$^{15}$ - 10$^{19}$ eV is larger than the volume
of the atmospheric layer, whose ability to convert downward
neutrino in observable air shower is negligible. Only horizontal
neutrino interaction in the atmosphere may be detected, at a much
lower rate than the HorTaus ones.} \label{Volume}
\end{figure}

The effective volume per unit surface is given by

\begin{eqnarray}
\frac{V_{Tot}(E_{\nu})}{A} = \frac{V_{Tot \, \oplus}(E_{\nu})}{ 2
\pi R^2_{\oplus}}= \nonumber  \\
\int^{\frac{\pi}{2}}_{0}
 \int^{D(\theta)}_{0} e^{-%
\frac{D(\theta) -x}{L_{\nu_{{CC}}}(E_\nu)}}  e^{\frac{-x}{%
l_{\tau}(E_{\tau})}}\sin{\theta}\cos{\theta}d\theta dx
\end{eqnarray}

where  $A$ is any arbitrary surface above the corresponding
effective volume. For instance this expression has been first
estimated for all the Earth. In this case A is just half of the
terrestrial surface, due to the request of selecting only the
upward direction.

Under the assumption that the $x$ depth is independent of $L_{\nu}$ and $%
l_{\tau}$, the above integral becomes:

\begin{eqnarray}
\frac{V_{Tot}(E_{\tau})}{A}=\left(\frac{l_\tau}{1-\frac{l_\tau}{L_\nu}}%
\right) \times \nonumber \\ \int^{\frac{\pi}{2}}_{0}
\left(e^{-\frac{D(\theta)}{L_{\nu CC }(\eta E_{\tau}))}}-
e^{-{\frac{D(\theta)}{l_\tau (E_{\tau})}}}\right)
\sin{\theta}%
\cos{\theta}d\theta \label{eq2}\end{eqnarray}

%

where the energy of the neutrino $E_{\nu}$ has been expressed as
a function
of $E_{\tau}$ via the introduction of the parameter $\eta = E_{\nu}/E_{\tau_f}$%
, the fraction of energy transferred from the neutrino to the
lepton. At energies greater than $10^{15}$ eV, when all mechanisms
of energy loss are neglected, $\eta = E_{\nu}/E_{\tau_f} =
E_{\nu}/E_{\tau_i} \simeq 1.2$, meaning that the 80\% of the energy
of the incoming neutrino is transferred to the newly born $\tau$
after the $\nu - N$ scattering (Gandhi 1996, 1998).

When the energy losses are taken into account, the final $\tau$
energy $E_{\tau_f}$ is a fraction of the one at its birth,
$E_{\tau_i}$. Their ratio $x_i = E_{\tau_f} / E_{\tau_i}$ is
related to $\eta$ by the following expression

\[
\eta(E_{\tau_f}) = \frac{E_{\nu}}{E_{\tau_f}} =
\frac{E_{\nu}}{E_{\tau_i}} \frac{E_{\tau_i}}{E_{\tau_f}} \simeq
\frac{1.2}{x_i(E_{\tau_f})} .
\]



This ratio defines the  fraction of energy of the outgoing $\tau$
compared to the incoming $\nu_{\tau}$.

 Once the effective volume is
found, we introduce an effective mass defined as

\begin{equation}
\frac{M_{Tot}}{A}=\rho_{out}\frac{V_{Tot}}{A} \label{eq_M_ltau}
\end{equation}

where $\rho_{out}$ is the density of the outer layer of the Earth crust: $\rho_{out}=1.02$ (water) and $%
2.65$ (rock).


Now we can neglect the severe energy degradation ($\eta \ll 1$) of
the HorTaus,  considering only those UHE $\nu_{\tau}$ that are
converted into $\tau$ within a smaller length $L_{\tau(\beta)}$
($L^{-1}_{\tau(\beta)} \propto \beta + (c \gamma_\tau t_\tau)^{-1} \leq
l_{\tau}$) (see Fargion et al. 2004 for detail). Note that $\beta$ is the
coefficient due to nuclear photoproduction and bremsstrahlung
energy losses.

Given that in general $e^{-{\frac{D(\theta)}{l_\tau}}} \ll
e^{-\frac{D(\theta)}{L_{\nu CC}}}$, the second exponential in the
integral in Eq. \ref{eq2} will be also neglected in writing the
following equations.

 The expression of the effective volume in this most
general case
becomes

\begin{equation}\begin{split}
\frac{V_{Tot}(E_{\tau})}{A}=\left(\frac{L_{\tau (\beta)}(E_\tau)}{1-\frac{%
L_{\tau (\beta)}(E_{\tau})}{L_{\nu_{CC}}(\eta E_\tau)}}\right) \times \\ 
\int^{\frac{\pi}{2}}_{0} 
e^{-\frac{D(\theta)}{L_{\nu CC
}(\eta E_{\tau}))}}
\sin{\theta}%
\cos{\theta}d\theta \label{Veff_lbeta}
\end{split}\end{equation}

where  the interaction length $L_{\tau (\beta)}$
(shown in Fig. \ref{l_tau} and compared to $l_{\tau }$) guarantees
a high energy outcoming $\tau$ even if outcoming from a thinner
Earth crust (see Fargion et al. 2004 for a more detailed discussion
of $L_{\tau (\beta)}$).

The terrestrial chord, $D(\theta ),$ shown in Fig. \ref{Dteta}, is
responsible for the $\nu _{\tau }$ opacity, and $L_{\nu }$ is the
interaction length for the incoming neutrino in a water equivalent
density, where $L_{\nu_{CC} }=(\sigma_{CC} \,n)^{-1}$. It should be
kept in mind that both $L_{\nu }$ and $D(\theta )$, converted into
the water equivalent
chord, depend on the number density $n$ (and the relative matter density $%
\rho _{r}$ of the inner shell tracks). In reality the Earth
interior has been idealized as an homogeneous sphere of water of
column depth $D(\theta$) (see Fig. \ref{Dteta}) and the \nutau
 interactions are considered for the water density.

We remind that the total neutrino cross section $\sigma _{\nu }$
consists of two main component, the charged current and neutral
current terms, but the $\tau $ production depends only on the
dominant charged current whose role will appear later in the event
rate number estimate. The interaction lengths $L_{\tau \beta }$,
$L_{\nu_{CC} },$ depends on the energy, but one should be careful
on the energy meaning. Here we consider an incoming neutrino with
energy $E_{\nu _{i}}$, a prompt $\tau $ with an energy $E_{\tau
_{i}}$ at its birth place, and a final outgoing $\tau $ escaping
from the Earth with energy $E_{\tau _{f}}$, after some energy
losses inside the crust. The final $\tau $ shower energy, which is
the only observable quantity, is nearly
corresponding to the latter value $E_{\tau _{f}}$ because of the negligible $%
\tau $ energy losses in air. However we must be able to infer
$E_{\tau _{i}}$ and the primary neutrino energy, $E_{\nu}$, to
perform our calculation. The effective volume  resulting from Eq.
\ref{Veff_lbeta} calculated for a detector with a 1 km$^2$
acceptance area is displayed in Fig. \ref{Volume}.





\section{Event Rate of Tau air showers for GZK neutrinos with EUSO and Auger}

After having introduced the effective volume we can estimate the
outcoming event number rate for EUSO for any given neutrino flux.
The consequent event rate for incoming neutrino fluxes may be
easily derived by:

\begin{equation}
\frac{dN_{ev}}{d\Omega dt}= \left( \int \frac{dN_{\nu}}{dE_{\nu}
d\Omega dA dt } \sigma_{N \nu} (E) dE \right) n \rho_r V_{Tot}
\end{equation}

\bigskip where $L_{\nu CC} = (\sigma_{N \nu} n)^{-1}$,
 $\Phi_\nu= \frac{d\,N_{\nu}}{d\,E_\nu}E_\nu=5 \cdot
10^{-18}\left( {\frac{E_\nu}{10^{19}eV}} \right)^{-\alpha + 1}
$cm$^{-2}$ sec$^{-1}$ sr$^{-1}$ describes as a flat spectrum
($\alpha = 2$) most of the GZK neutrino flux and as a linearly
increasing spectrum ($\alpha = 1$)  the Z-burst model; $\rho_r$ is
the density of the most external layer (either rock or water). The
assumption on the flux may be changed at will  and the event number
will scale linearly according to the model.

\begin{figure}[t]
\includegraphics[angle=270,scale=0.3]{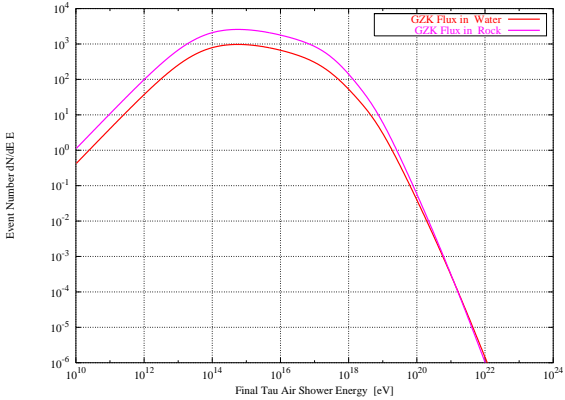}
\caption{Number of EUSO Event for HorTaus in 3 years record as a
function of the outgoing lepton tau ($L_{\tau (\beta)}$ as
interaction length), including the finite extension of the
horizontal atmospheric layer. At energy $E_{\tau} = 10^{19}$ eV,
the event number is $N_{ev}= 3.0$ ($\protect\phi_{\protect\nu}
E_{\nu} / 50$ eV cm$^{-2}$ s$^{-1}$ sr$^{-1}$) for the
water and $N_{ev}= 6.0$ ($\phi_{\protect\nu} E_{\nu} / 50$ eV cm$^{-2}$ s$%
^{-1}$ sr$^{-1}$) for the rock. The resulting number of events has
been calculated for an initial GZK neutrino flux: $\propto E^{-2}$
.} \label{EUSOevent}
\end{figure}

In Fig. \ref{EUSOevent} we show the expected number of event for
EUSO 
where we have included the Earth's atmosphere and we have used
$L_{\tau (\beta)}$, so that we may express the results as a
function of the final $\tau$ lepton energy.

%
%
%

As one can see from Fig. \ref{EUSOevent}, at energy $E = 10^{19}$
eV the general expected event rate is given by:

\[ N_{ev}\,= 5\cdot 10^{-18}cm^{-2}s^{-1}sr^{-1}\,
\left({\frac{V_{eff} \rho_r}{L_{\nu CC}}}\right) (2 \pi\,\eta_{Euso} \Delta t)  \]  
\[ \times \left( \frac{\Phi_{\nu} E_{\nu}}{50 eV cm^{-2} s^{-1}
sr^{-1}} \right) \eta^{-\alpha} \left( \frac{
E_{\tau}}{10^{19} \,  eV} \right)^{- \alpha + 1}\] \\
%

where $\eta_{Euso}$ is the duty cycle fraction of EUSO,
$\eta_{Euso} \simeq 10\%$, $\Delta \,t\ \simeq 3$ $years$.
%
%
 Such number of events greatly exceed previous results by at least two orders of magnitude (Bottai et al. 2003)
 but are comparable, but below more recent estimates (Yoshida et al. 2004).

\begin{figure}[t] 
\includegraphics[width=80mm,height=45mm,angle=0]{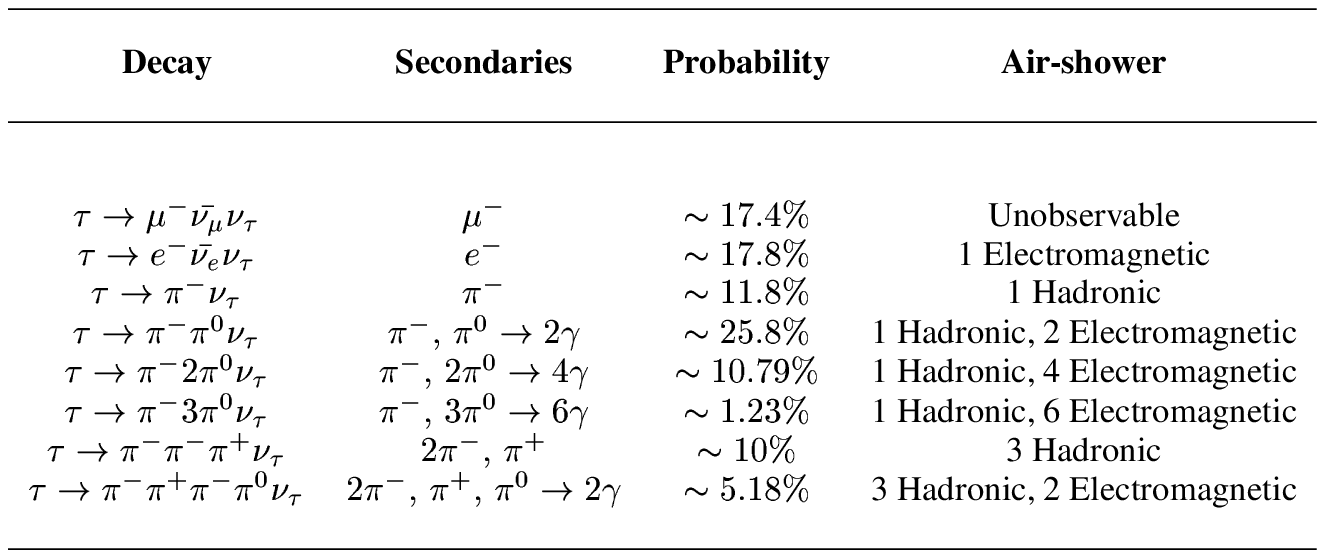}
\caption{Tau decay channels.The probabilities for pion
secondaries take into account the contribution of  $K$ mesons
originated from $\tau$ decay \cite{Fargion 2002a}} \label{Tab_tau}
\end{figure}

We have also calculated the number of event rate for the Auger
detector (Fargion et al. 2004). The number of events is slightly
above the unity at $E = 10^{19}$ eV (see Fig.19 in Fargion et al.
2004) for scintillator detection and near 0.1 events/yr for
photoluminescent detectors. However at energy as low as $E =
10^{18}$ eV in GZK model the number of events possibly detectable
by the scintillators of Auger increases to $26.3$ for a rock
layer (for scintillator detectors) and 2.6 events for photofluorecence ones: the most remarkable signature will be a strong azimuthal
asymmetry (East-West) toward the high Andes mountain chain. The
Andes shield UHECR (toward West) suppressing their horizontal
flux. These horizontal showers originated in the atmosphere at
hundreds of km, are muon-rich as well as poor in electron pair
and gammas and are characterized by a short time of arrival. On
the other hand the presence of the mountain ranges at $50-100$ km
from each of the Auger detector will because of the mountain geometry by a numerical
factor $2-6$ (for each given geographical configuration) the
event number respect to the  East direction. These EeVs rare nearby
(few or tens of km) tau air-shower will be originated by
Horizontal UHE neutrinos interacting inside the Andes. These
showers will have a large electron pair, and gamma component
greater than the muonic one, and a dilution in the time of
arrival (microseconds over nanoseconds in horizontal $\mu^\pm$ by
UHECR at the horizon). Timing, spectra, muon-gamma ration and directionality will clearly mark their $\nu_\tau$ nature.

\section{The differential rate of Tau air showers}

%



We introduce now a differential expression of the number of events
which allows to calculate the number of events as a function of the
angle $\theta$. We can rewrite the expression of the effective
volume given in Eq. \ref{Veff_lbeta} as a differential volume for
each arrival angle $\theta$:

\begin{equation}\begin{split}
\frac{dV}{ d\theta\, d\phi \, d\Omega \, dA}= \left[
1-e^{\frac{-L_{0}}{c\,\gamma_\tau \tau}}\right]
\times \\
 {l_\tau (E_\tau)}
\frac{
e^{-\frac{D(\theta)}{L_{\nu
CC}}}
}{\left( 1-\frac{l_\tau
(E_\tau)}{L_{\nu CC}}\right)}\sin\theta \, \cos\theta
\end{split}\end{equation}

and we obtain the following expression for the differential rate of
events

\begin{equation}\begin{split}
\frac{dN_{ev} E}{dE d \Omega \, d\theta\, d\phi\, dt\, dA}=
\Phi_{\nu_0} \eta^{-\alpha} \left(
\frac{E_\tau}{E_{\nu_0}}\right)^{-\alpha + 1} \rho_r \times
\\
\left[ 1-e^{\frac{-L_{0}}{c\,\gamma_\tau \tau}}\right]
 \frac{l_\tau (E_\tau)}{L_{\nu CC}}
\frac{e^{-\frac{D(\theta)}{L_{\nu
CC}}}
}{\left( 1-\frac{l_\tau (E_\tau)}{L_{\nu CC}}\right)}\sin\theta \,
\cos\theta
\end{split}\end{equation}

with $\eta=1.2$ and $E_{\nu_0} = 10^{19}$  eV.

If we now integrate on the solid angle $d\Omega$ (half side) we
obtain the above formula multiplied by a factor $2\pi$.

%

\begin{figure}[t] 
\includegraphics[width=80mm,height=55mm,angle=0]{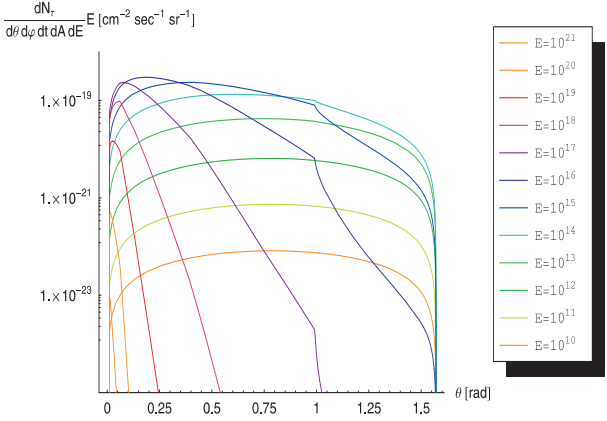}
\caption{
 The differential number of event rate of $\tau$
leptons (HorTaus) for an input GZK neutrino flux. As in previous
Figures we are assuming that $\tau$'s are escaping from an Earth
outer layer made of rock. Note the discontinuity at $\theta \simeq
1$ rad, due
to the corresponding  inner terrestrial  higher density core (see Fig. \ref{Dteta}). 
} \label{tau_mu}
\end{figure}

\begin{figure}[t] 
\includegraphics[width=80mm,height=55mm,angle=0]{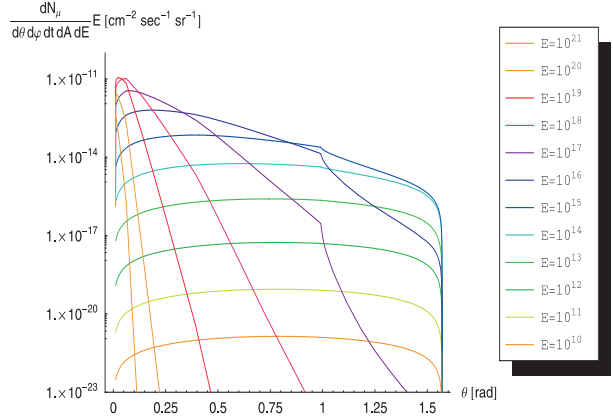}
\caption{
The differential number of event rate of the secondary muons
produced by the decay in flight of $\tau$ leptons in the Earth's
atmosphere: HorTaus. As in previous Figures we are assuming an
input GZK neutrino flux and that $\tau$'s are escaping from an
Earth outer layer made of rock.} \label{tau_mu}
\end{figure}



\begin{figure}[t] 
\includegraphics[width=80mm,height=55mm,angle=0]{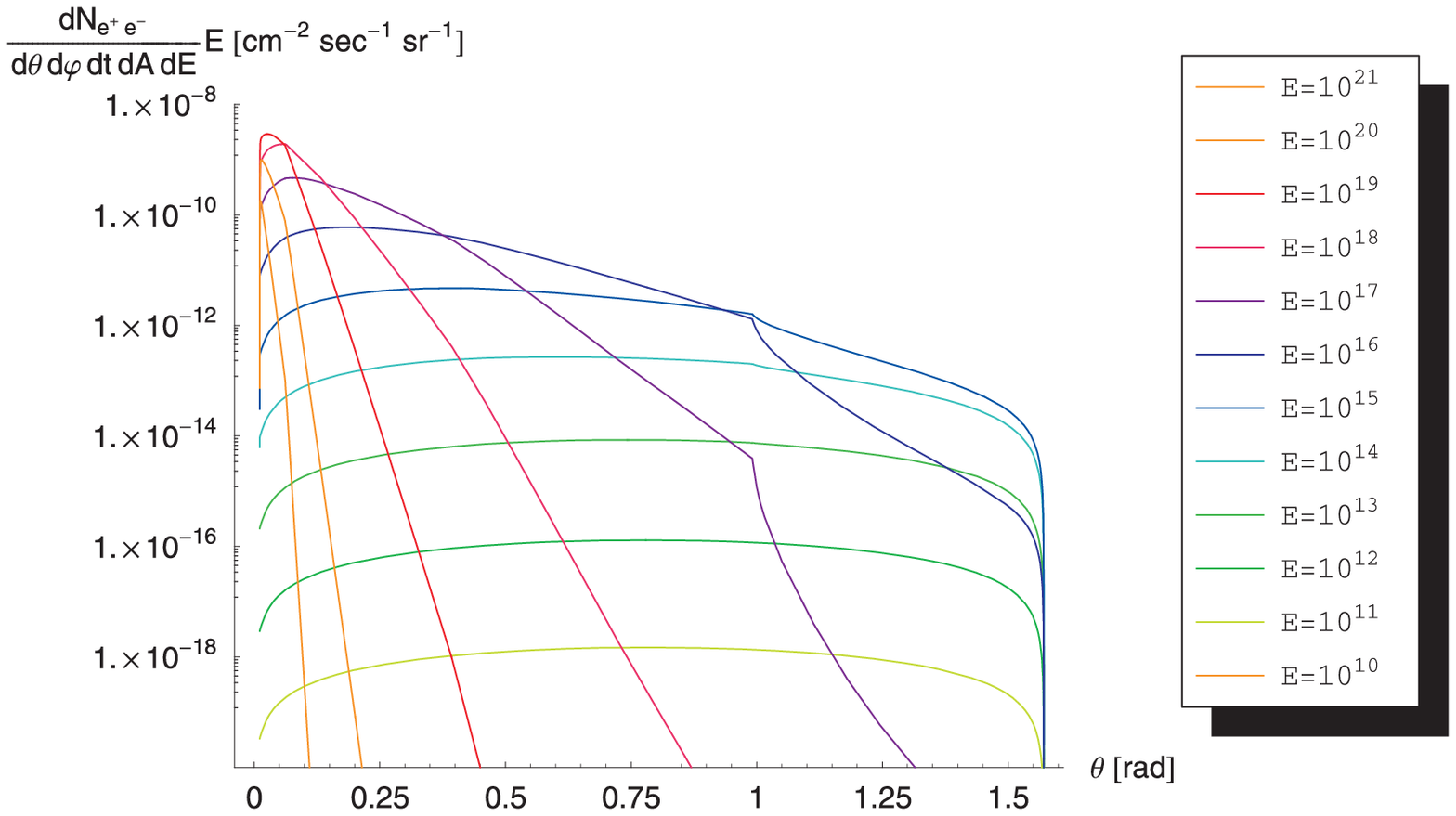}
\caption{
The differential number of event rate of secondary electron pairs
originated from the decay of $\tau$ leptons for an input GZK
neutrino flux. As in previous Figures we are assuming  that
$\tau$'s are escaping from an
Earth outer layer made of rock.
}\label{e_gamma}
\end{figure}

\begin{figure}[t] 
\includegraphics[width=80mm,height=55mm,angle=0]{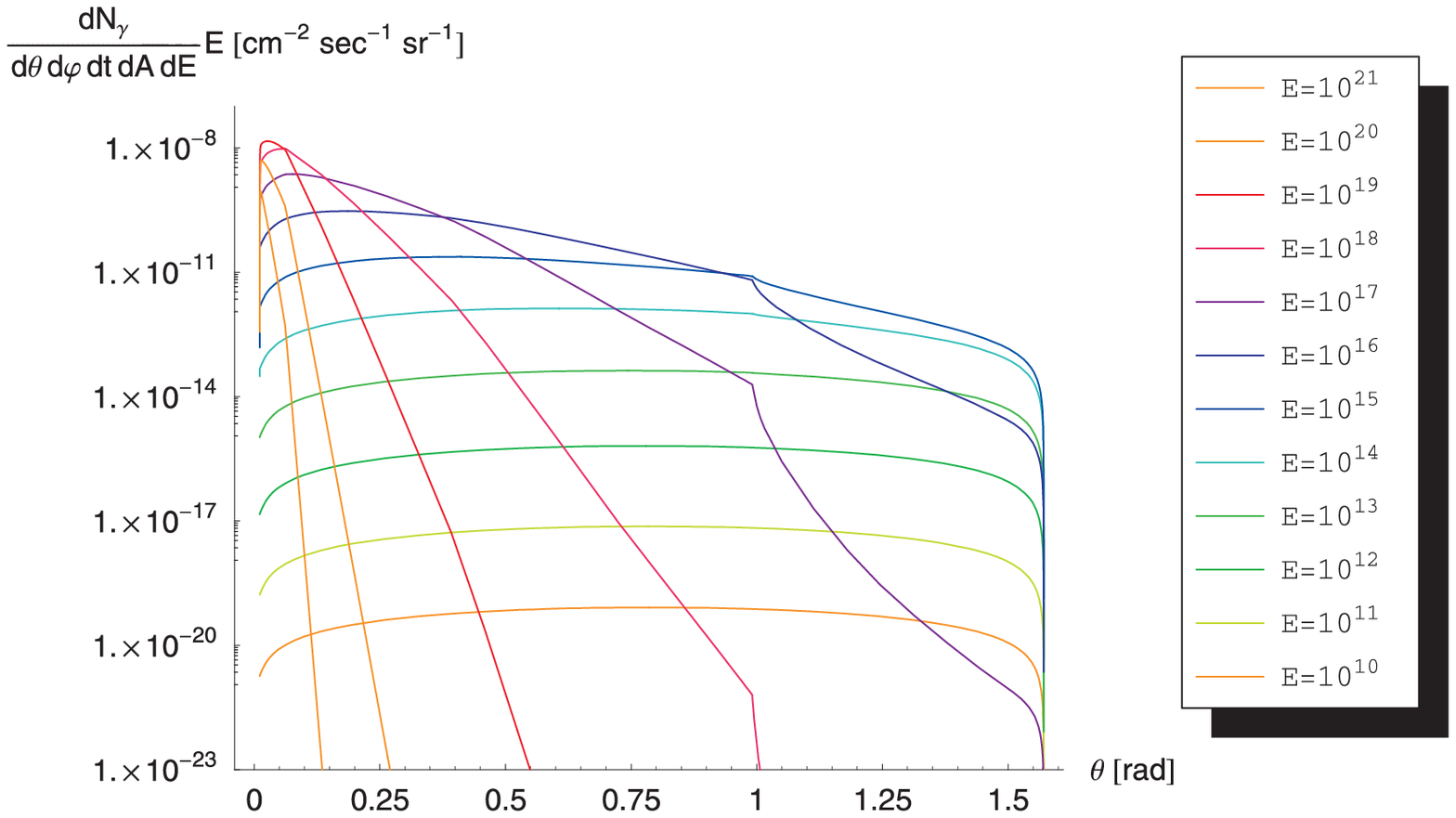}
\caption{
The differential number of event rate of secondary $\gamma$'s
originated from the decay of $\tau$ leptons for an input GZK
neutrino flux. As in previous Figures we are assuming  that
$\tau$'s are escaping from an Earth outer layer made of rock.}
\label{e_gamma}
\end{figure}

Given the $\tau$ number of events we can calculate the rates of
$\mu$, $e^\pm$ pairs and $\gamma$, which originates as secondary
particles from the $\tau$ decay (see Fig. \ref{Tab_tau}). The
number of muons is related to the total number of decaying pions
and according to Matthews (2001)  is given by


\begin{equation}
N_\mu \simeq 3\cdot 10^5 \left( \frac{E_\tau}{PeV}\right)^{0.85}
\end{equation}

\begin{equation}
N_{e^+ e^-} \simeq 2\cdot 10^7 \left( \frac{E_\tau}{PeV}\right)
\end{equation}

\begin{equation}
N_{\gamma} \simeq  10^8 \left( \frac{E_\tau}{PeV}\right)
\end{equation}

and we obtain finally

\begin{equation}
\frac{dN_{i\,ev}}{d\theta\, d\phi\, dt\, dA}(E, \theta)=  N_i \cdot
\frac{dN_{\tau \,ev}\,E_\tau}{ dE \, d\theta\, d\phi\, dt\,
dA}\nonumber \label{Nevent_multiplicity}
\end{equation}


We show in Fig. \ref{tau_mu} the average differential rate of
$\tau$'s and the secondary $\mu^\pm$, $e^{\pm}$ and $\gamma$
bundles from the decay of $\tau$ leptons.
  We find
that the muon signal at the horizon, related to Earth skimming tau
neutrinos is above 10$^{-12}$ - 10$^{-11}$ cm$^{-2}$ s$^{-1}$
sr$^{-1}$. One should notice that the  muonic background   produced
by atmospheric neutrinos (CR $\rightarrow \mu^{\pm} \rightarrow
\nu_{atm} \rightarrow \mu^{\pm}$) below the horizon approaches at beast
to the value $\Phi_{\mu_{atm}}\simeq 2\cdot 10^{-13}$ cm$^{-2}$ s$^{-1}$ sr$^{-1}$, which is at
least one order of magnitude lower than what we have obtained from
our calculation for minimal GZK $\nu_{\tau}$ fluxes (see Fig.
\ref{tau_mu}).

\begin{figure}[t] 
\includegraphics[width=80mm,height=65mm,angle=0]{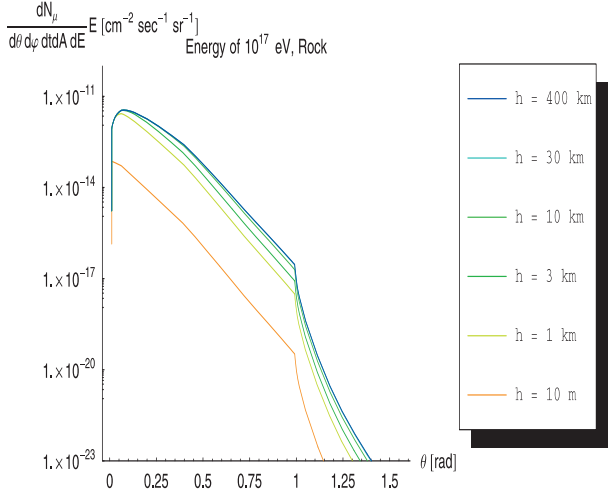}
\caption{The average differential event rate of secondary muon
leptons  from HorTaus having assumed $E_{\tau} = 10^{17}$ eV and an
input GZK-like neutrino flux at different altitudes. Again we are
assuming that $\tau$'s are escaping from an Earth outer layer made
of rock. Note the discontinuity of the angular spectrum due to the
inner core of the Earth at $\theta \simeq 1$ rad and the asymptotic
behaviour for $h \gg 1$ km.} \label{MU_FLUX17}
\end{figure}

\begin{figure}[t] 
\includegraphics[width=80mm,height=65mm,angle=0]{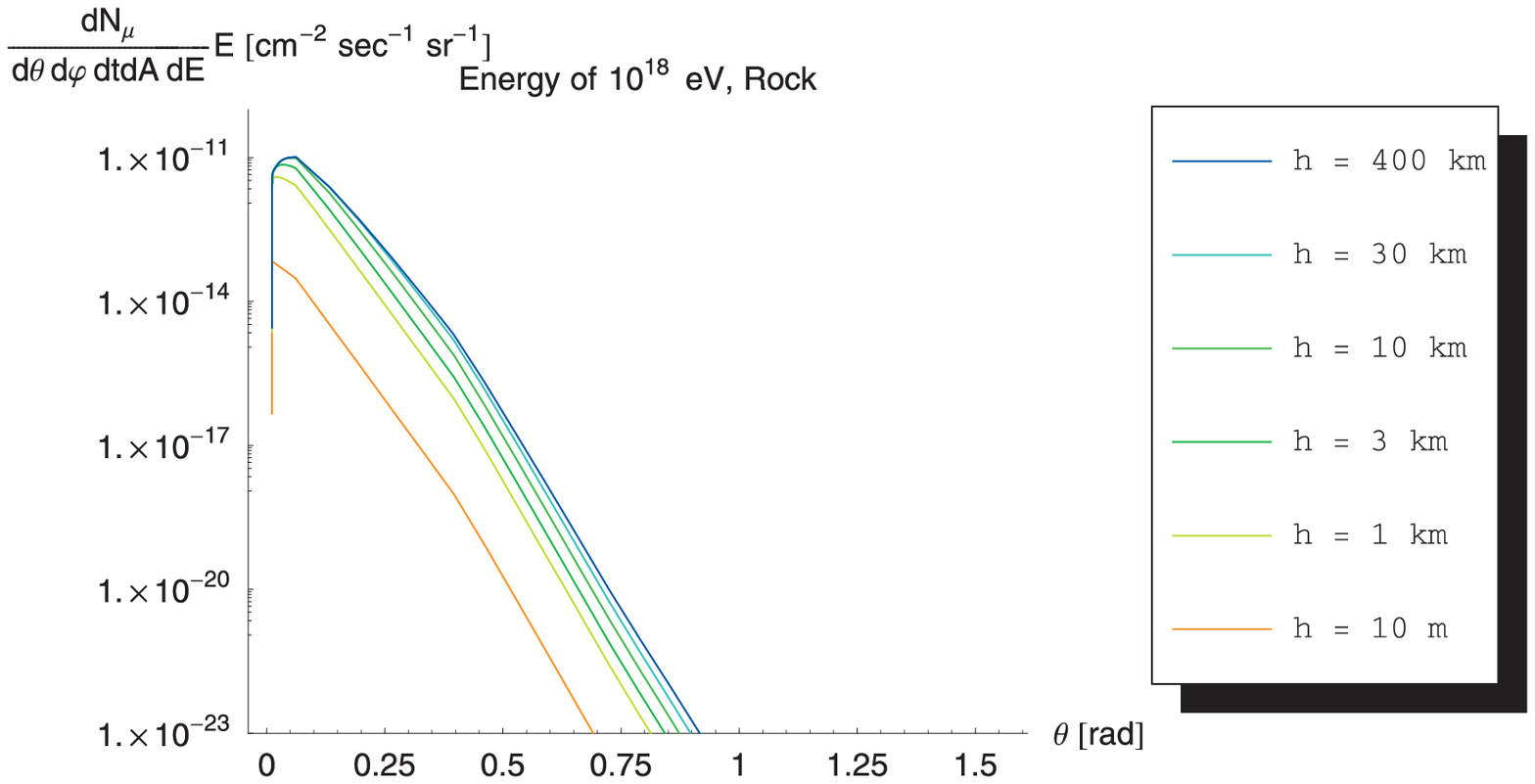}
\caption{The differential event rate of muon leptons having assumed
$E_{\tau} = 10^{18}$ eV and an input GZK-like neutrino flux. Again
we are assuming that $\tau$'s are escaping from an Earth outer
layer made of rock. Note that because the Earth is nearly opaque to
EeV \nutau, only Earth-skimming neutrinos nearly horizontally are
visible. The discontinuity in the spectrum at $\theta \sim$ 1 rad
occurs also at energies as large as 10$^{18}$ eV, but at lowest
flux, and we do not see it in this picture.} \label{MU_FLUX18}
\end{figure}

Moreover this muonic shower would have a significant $\gamma$
component, with a high number of  photons - $N_{\gamma}/N_{\mu}
\sim 10^2$ (Cillis \& Sciutto 2001; see also Fig. 15 in Cronin
2004)
 - because of the $\tau$ decay channels into both charged and neutral
pions as one can see from Fig. \ref{Tab_tau}.

On the other hand, horizontal UHECRs will not represent a source of
contamination for our signal because the horizontal $\gamma$'s
produced in the hadronic shower inside the atmosphere would be
exponentially suppressed at large slant depth ($X_{max} > 3000$ g
cm$^{-2}$) and large zenith angles ($\theta > 70^\circ$) (Cillis \&
Sciutto 2001). Only $\mu^\pm$ can survive when propagating through
the atmosphere at large zenith angles. Such muons would also be
source of parasite $\gamma$ signal - due to the $e^\pm$ pair produced in
the $\mu^\pm$ decay in flight - but the gamma-to-muon-number ratio
would be now approximately $\lesssim$ 1.

Therefore, this difference would allow to distinguish gamma-rich HorTaus from common
horizonal gamma-poor UHECR events to reveal  UHE earth-skimming $\nu_{\tau}$'s
.

For a more precise approach to the calculation of the rate of
events one has also to take into account that the number of events
varies as a function of the height $h$ at which the observer is
located to detect the muonic, electronic and gamma shower. Therefore
we
introduce 
an additional factor

$$\frac{dN_{\mu\,ev}}{  d\theta\, d\phi\, dt\, dA}(E, \theta,
h)=$$
\begin{equation}
\left( 1 - e^{-\frac{h}{h_s}} \right) \frac{dN_{\mu\,ev}}{
d\theta\, d\phi\, dt\, dA}(E, \theta)
\end{equation}

where the parameter $h_s$ that we have introduced as

\begin{equation}
h_s=R_{\tau} (E_{\tau}) \sin{\theta} +
\frac{X_{Max}(E_\tau)}{\rho_r}\sin{\theta}
\end{equation}


defines the optimal height where the shower can reach its maximal
extension at the corresponding energy $E_{\tau}$.  This is the sum
of the height reached by the $\tau$ in the atmosphere before its
decay ($R_{\tau} \sin \theta$), where we have neglected $\tau$
energy losses in the atmosphere, and the altitude reached by the
secondary particles of the shower which is related to  the
parameter $X_{max}$. Note that $R_{\tau} = 4.9 (E_\tau/10^{17} \,
eV)$ km and $X_{max}/\rho_r = 5.7 + 0.46 \ln (E_{\tau}/10^{17} eV)$
km. Here we have considered the air density $\rho_r = 1.25 \times
10^{-3}$ g cm$^{-3}$ constant and equal to the value at the sea
level. At low energies ($10^{15}$ - $10^{16}$ eV) the second term
is dominant and $\rho_r$ can be considered as constant because the
$\tau$ lepton travels for less than 1 km before it decays. At
higher energy ($10^{17}$ - $10^{19}$ eV) the first term is
dominant, and we can neglect the way the exact value of  $\rho_r$
changes with the altitude. We show our results in Fig.
\ref{MU_FLUX17} and Fig. \ref{MU_FLUX18} where the secondary $\mu^\pm$ fluxes are described for two given $E_\tau$ energies.

\section{Conclusions}

Horizontal showers from normal hadrons (or gammas) are strongly
depleted of their electromagnetic component because of the large
slant depth ($X_{max} > 3 \times 10^4$ g cm$^{-2}$), while
horizontal tau air showers are not. Indeed "young" HorTaus either
of hadronic (67\%) or electromagnetic (33\%) nature (see Fig.
\ref{Tab_tau}) at their peak shower activity are expected  to have
a large $N_{\gamma}/N_{\mu}$ ratio, greater than $10^2$ (but with
a characteristic energy ratio $E_{\gamma}/E_{\mu} < 10^2$). Old
horizontal showers would have $N_{\gamma}/N_{\mu} \simeq 1$. This
difference would allow to distinguish and disentangle HorTaus from
horizontal UHECR events, even in absence of good angular
resolution, opening a new perspective in the UHE neutrino
Astronomy. The secondary fluxes of muons and gamma bundles made
by incoming GZK neutrino fluxes and their HorTau showers, is well
above the noise (by one-two order of magnitude) made by up-going
muons, Earth-Skimming trace of atmospheric neutrinos. The neutrino
signals at energies much above EeV  may be even better probing the
expected harder neutrino Z-Burst model spectra
\cite{Fargion-Mele-Salis99}, \cite{Weiler99}.

The peak fluences we find  in the $\mu$ and $\gamma$ component at
the horizon ($\pm 5^\circ$) will give a signal well above the
background produced by atmospheric  $\nu$'s. We did not discussed
the albedo muons whose fluxes measured  by Nemo-Decor experiments,
$\phi_{\mu} \lesssim 10^{-9}$ cm$^{-2}$ s$^{-1}$ sr$^{-1}$, are
made mostly by single tracks \cite{decor}. Because pair (or
triple) bundle muons are much
 rarer ($\phi_{\mu_{pair}} < 10^{-4} \phi_{\mu_{single}} \approx 10^{-13}$ cm$^{-2}$ s$^{-1}$
 sr$^{-1}$), the search and detection of muon bundles by GZK
 HorTaus at a minimal rate of $10^{-12}$ cm$^{-2}$ s$^{-1}$
 sr$^{-1}$ (over an area of $10^2$ m$^2$) will lead, in a year, to
 about 30 muons possibly clustered in five - ten multiple bundles.
 These events will be reinforced by hundreds or thousands of
 associated collinear gamma flashes.
A detector with an area of few tens or hundreds of square meters
pointing to the horizon from the top of a mountain would be able
to reveal the GZK $\nu_{\tau} - \tau$ young showers (Iori, Sergi
\& Fargion 2004). The characteristics of a prototype twin
crown-like array detector to be placed on mountains, balloons, or
satellites \cite{Fargion2001a} will be discussed in detail
elsewhere. The simultaneous sharp $\gamma$ bundle at
$\phi_{\gamma} \sim 10^{-9} \div 10^{-11} $ cm$^{-2}$ s$^{-1}$
sr$^{-1}$ and the "burst" of electron pair at $\phi_{e^+e^-} \sim
10^{-9} $ cm$^{-2}$ s$^{-1}$ sr$^{-1}$ would give evidence of
unequivocal $\tau$ signature. It should be remind that the
neutrino interaction enhancement by TeV new Physics (as shown in
Fig$1$ and Appendix C) \cite{Fargion 2002a}, would produce also an
increase of hundreds or thousands time in HorTaus beyond a
mountain Chain (like Auger) than standard weak interactions would
do. Therefore Auger must soon detect either the Ande Shadows for
$old$ UHECR (at zenith angle larger than $85^o -88^o$) by their
absence as well as  $young$ \cite{Fargion1999}, \cite{Bertou2002}
HorTaus productions by the mountain themselves due to New Physics
at TeV. In a few years Auger might be even able to observe also GZK
neutrinos induced HorTaus at EeV energies (by sure if some
technical improvement will be made). EUSO experiment will be able
to see at least  half a dozen of events of HorTaus mainly
enhanced along the Continental Shelves or Mountain edges.
 To conclude we want to remind that inclined-vertical PeVs
$\tau$ air showers (UpTaus) would nearly always be source of
$\gamma$ "burst" surviving the atmosphere opacity. These sharp
UpTaus (with their companion HorTaus above and near EeVs) might be
observed by satellites as brief Terrestrial Gamma Flashes (TGF).
Indeed we identified a possible trace of such events in BATSE
record (taken during the last decade) of $78$ upgoing TGF
maybe associated with galactic and extragalactic UHE neutrino
sources \cite{Fargion 2002a}.


\begin{thebibliography}{4}





%
%
%
%
%
%
%
%
\bibitem{Bertou2002}  {\normalsize  Bertou, X.,
Billoir, P., Deligny, O., Lachaud, C., Letessier, S.A,  2002,
Astropart. Phys.,  17,  183}
%
\bibitem{Bottai2002}{\normalsize Bottai, S., Giurgola, S., 2003, BG03, Astrop. Phys., 18, 539}
%
\bibitem{Buga2003}{\normalsize Bugaev, E.V., Montaruli, T., Sokalski, I.A., 2004, Phys. Atom. Nucl., 67, 1177; astro-ph/0311086}

\bibitem{Cillis2001}{\normalsize Cillis, A.N.,  \& Sciutto, S.J., 2001, Phys. Rev. D64, 013010}


\bibitem{Cronin2004}{\normalsize Cronin, J.W., 2004, TauP Proceedings, Seattle 2003; astro-ph/0402487}


%
%
%
%
%
%
\bibitem{Fargion-Mele-Salis99}
{\normalsize Fargion, D., Mele, B., \& Salis, A., 1999, ApJ 517,
725; astro-ph/9710029 }
%
\bibitem{Fargion1999}  {\normalsize
Fargion, D., Aiello, A., Conversano, R.,  1999, 26th ICRC HE6.1.10
396-398; astro-ph/9906450 }
%

\bibitem{Fargion 2002a}  {\normalsize Fargion, D.,
2002, ApJ, 570, 909; astro-ph/0002453; astro-ph/9704205 }

\bibitem{Fargion2004}{\normalsize Fargion, D., De Sanctis Lucentini, P.G., De Santis, M., Grossi, M.,
 2004, ApJ, 613, 1285; hep-ph/0305128.}
%
%
\bibitem{Fargion2001a}{\normalsize Fargion, D., 2001, Proc. 27th
ICRC (Hambourg) HE2.5 1297-1300; astro-ph/0106239 }
%
%
%
%
%
%
%
%
%
%
%
%
%
\bibitem{Feng2002}  {\normalsize Feng, J.L.,
Fisher, P., Wilczek, F., \& Yu, T.M., 2002, Phys. Rev. Lett. 88,
161102; hep-ph/0105067 }

%
\bibitem{Gandhi96}{\normalsize Gandhi, R., Quigg, C., Reno, M.H., Sarcevic, I., 1996, Astrop. Phys., 5,  81}
%
\bibitem{Gandhi98}{\normalsize Gandhi, R., Quigg, C., Reno, M.H., Sarcevic, I., 1998, Phys. Rev. D., 58, 093009}
%

\bibitem{Iori04}{\normalsize Iori, M., Sergi, A., Fargion, D., 2004, astro-ph/0409159}


\bibitem{Jones04}{\normalsize Jones, J., Mocioni, I., Reno, M.H., Sarcevic, I., 2004  Phys. Rev. D
 , 69, 033004}
%
%
%
%
%

\bibitem{Matthews01}{\normalsize J.C. Matthews (2001), Proc. 27th ICRC (Hamburg), 1, 161.}

%

\bibitem{Tseng03}{\normalsize Tseng, J.J, Yeh, T.W., Athar, H., Huang, M.A., Lee, F.F., Lin, G.L., 2003, Phys.
Rev. D68, 063003}
%
%
%
\bibitem{decor} I.I. Yashin Proceeding IV Int. Conf. NANP'03 23 June 2003 Dubna


\bibitem{Yoshida2004}{\normalsize  Yoshida, S.,  Ishibashi, R., Miyamoto, H., 2004, Phys. Rev. D69, 103004;
astro-ph/0312078}
%

\bibitem{WB97}{\normalsize  Waxman, E., Bachall, J., 1997, Phys. Rev. Lett., 78, 2292}

\bibitem{Weiler99}  {\normalsize  Weiler, T.J., 1999, Astropart. Phys. 11, 303}


%

\end{thebibliography}
\end{document}